# Spin transfer driven resonant expulsion of a magnetic vortex core for efficient rf detector

S. Menshawy[1,2,a], A.S. Jenkins[2,4], K.J. Merazzo[3], L. Vila[3], R. Ferreira[4], M.-C. Cyrille[3],

U. Ebels[3], P. Bortolotti[2], J. Kermorvant[1], V. Cros[2]

[1]*Thales Communications and Security, 4 avenue des Louvresses 92230 Gennevilliers, France*

[2] *Unité Mixte de Physique CNRS, Thales, Univ. Paris-Sud, Université Paris-Saclay, 91767 Palaiseau, France*

[3]*SPINTEC, Univ. Grenoble Alpes / CEA / CNRS, 38000 Grenoble, France*

[4]*International Iberian Nanotechnology Laboratory (INL), 4715-31 Braga, Portugal*

[a]email : samh.menshawy@thalesgroup.com

Spin transfer magnetization dynamics have led to considerable advances in Spintronics, including opportunities for new nanoscale radiofrequency devices. Among the new functionalities is the radiofrequency (rf) detection using the spin diode rectification effect in spin torque nano-oscillators (STNOs). In this study, we focus on a new phenomenon, the resonant expulsion of a magnetic vortex in STNOs. This effect is observed when the excitation vortex radius, due to spin torques associated to rf currents, becomes larger than the actual radius of the STNO. This vortex expulsion is leading to a sharp variation of the voltage at the resonant frequency. Here we show that the detected frequency can be tuned by different parameters; furthermore, a simultaneous detection of different rf signals can be achieved by real time measurements with several STNOs having different diameters. This result constitutes a first proof-of-principle towards the development of a new kind of nanoscale rf threshold detector.

Spintronic devices provide new functionalities that go beyond those possible with conventional electronic devices. Among them are the remarkable radiofrequency (rf) properties associated to spin torque driven magnetization dynamics [1]. Indeed, the increasing of performances of spin transfer nano-oscillators (STNO) and their diversification (uniform magnetic state, vortex, double vortices…) [2, 3, 4, 5, 6] allows considering different kind of spintronic devices; the one of interest in this paper is the detection of an rf signal. Beyond their ability to emit a rf signal, this approach for rf detection arises from the injection of an rf current through the STNO device that in turn generates a dc voltage across the STNO device when the frequency of the rf current is equal to the eigenfrequency of the modes that is considered, being in our case the gyrotropic vortex mode [7, 8, 9]. This rectification dc voltage is dubbed as the *spin-diode effect* in reference to conventional semiconductor diodes. Indeed, because the resonant frequency can be tuned, for example using an external magnetic field, it was proposed to use this spin torque effect as a basis for a nanoscale radiofrequency detector [10, 11]. In the perspective of developments, one of the most important issues remains to increase the sensitivity of the radiofrequency detection. In fact, the maximum generated voltage for an STNO with uniform magnetization has been found to be about 12 000 V/W [11, 12]. The wealth of the spin torque magnetization dynamics, and notably for a vortex configuration, has recently allowed us to establish a new effect that is the spin torque driven resonant expulsion of a vortex [13]. This behavior, observed for large rf amplitude, results into a large change of voltage across the STNO device as soon as a finite dc current is applied. Converting the detected change of voltage (about 2 mV in [13]) when the vortex is expelled for an external frequency equals the gyrotropic frequency, we find a considerable increase of the effective sensitivity (40 000 V/W).

The purpose of the present work is to further investigate this new phenomenon in the perspective of the development of a new generation of nanoscale rf threshold detector.

Measurements have been performed on circular MTJs nanopillars with diameters of 200 nm, 300 nm, 400 nm and 500 nm respectively. The stack is composed of a Synthetic antiferromagnet (SAF) / MgO(1 nm) / NiFe(7 nm). The pinned SAF layer is a PtMn(15 nm) / CoFe(2.5 nm) / Ru(0.85 nm) / CoFeB (3 nm) multilayer. The NiFe layer (7 nm) is the free layer with magnetic vortex as the ground state. Standard dc and high frequency current sources are used to source $I_{dc}$ and $I_{rf}$ respectively and the voltage is measured with a digital voltmeter connected via a bias-tee. The external magnetic field is applied with an electromagnet radiating a field between -0.85 mT and 0.85 mT. For measurements including several STNOs in parallel, a dedicated PCB circuit has been fabricated allowing to distribute the incoming rf current over the tested STNOs thanks to a power divider (with a loss of - 6 dBm) and to measure simultaneously the voltage across each STNOs using a digital voltmeter connected via a bias tee. All measurements have been performed at room temperature.

As explained before, the vortex-based STNO have two possible responses depending on the power amplitude of the rf current injected. For a small rf current (typically under $I_{rf}$ = 0.4 mA), a spin-diode effect is observed when the external frequency leads to a resonant excitation of the vortex core, as shown in Fig.1a). In this case, the mixing between the oscillating current and the time varying resistance generated by spin transfer vortex dynamics generates a rectification voltage. For a large rf current (typically above $I_{rf}$ = 0.4 mA for this STNO), the oscillation radius of the vortex core induced by excitation associated to an rf current at the resonant frequency becomes too rapidly larger than the physical diameter of the STNO meaning that the vortex has been expulsed. In this regime, the voltage generated by vortex oscillators can be described as $\Delta V = \Delta R_{exp} I_{dc} + I_{rf} \Delta R_{osc} \cos(\omega_s t)$, where the first part is the dc voltage associated with the change in resistance, $\Delta R_{exp}$, which occurs when the core is expelled and the second part is the conventional spin-diode effect due to the variable resistance, $\Delta R_{osc}$. In Fig.1b) we display a typical measurement recorded in the vortex core expulsion regime. When the frequency of the rf current injected through the structure approach the resonant frequency, the magnetic configuration of the free layer of the STNO goes from a vortex configuration to a quasi-uniform magnetic state. This is accompanied by a sudden and significant change of the resistance due to the magnetoresistive properties of the STNO. When a non-zero dc current is applied, this results in a large voltage variation across the device. Eventually, when the external frequency is beyond the resonant frequency, the free layer of the STNO returns at its more stable configuration, i.e., the vortex core is re-nucleated.

One of the advantages of this new type of rf threshold detectors is that the detected frequency can be tuned by several approaches, which is a key aspect from a technological point of view. For example, in Fig 2a) we show the voltage signals measured from three different STNOs having three different diameters, i.e., 300, 400 and 500 nm. Note that these measurements have been recorded separately. The objective here is to show that several frequencies can be detected simultaneously providing a first proof-of-concept of a real-time nanoscale spectrum analyzer. Hence, simply by changing the dimensions of STNOs in an array, it would be feasible to target a range of frequencies to be detected ranging from about a few tens of MHz up to more than 1.5 GHz, that is practically important for some targeted telecommunication devices. Furthermore, an additional asset of the vortex-based rf detector is that the frequency that can be detected is tunable through the application of an external in-plane magnetic field. This external field (that could be locally created by a current line, to remain compatible with a large integration level of this device) tends to shift the vortex core position from the center of the STNO. Note that in the experiments presented in Fig. 2 d), the in-plane field component is achieved by tilting slightly the STNO with respect to the normal direction (the x-axis shows the amplitude of the in-plane field). Another interesting feature shown in Fig. 2 d) is that, even if the central frequency between expulsion and re-nucleation (resp. blue and red points in Fig. 2 d)) varies when the field amplitude changes, the range of frequency in which the vortex is expulsed, i.e., between the blue and the red points in Fig. 2d), remains almost constant with field. In other words, the detector bandwidth is almost not changed whereas the detected frequency can be tuned. Finally, we have also studied how the effects of vortex expulsion and re-nucleation are affected when the power amplitude is changed, see Fig. 2 c). Whereas the frequency at which the expulsion occurs is decreasing when the power increases, which is consistent with the linear increase of the spin-torque driving force with the amplitude of the rf current, the frequency of re-nucleation (red points in Fig. 2c)) remains constant. This latter behavior indicates that the mechanism leading to the re-nucleation of the vortex in the STNO is probably not a resonant phenomenon, at least under the conditions of dc current and external field that are used. Interestingly, in our previous study by Jenkins et al. [8] in very similar devices, that has been done with a dc current much lower than the threshold current for auto-oscillations, we found that the frequency of re-nucleation was also modified by the increase of the rf power. Further investigations are in progress to get more insights about these different observations obtained at distinct ranges of rf power amplitude. This is also a key point to understand in view of the targeted applications that should cover a large spectrum of rf amplitude to be detected.

A further step towards the development of an instantaneous threshold rf detector has been to demonstrate the simultaneous detection of several rf signals. This objective can be achieved by connecting several STNOs in parallel. As demonstrated in Fig 3, we succeed to observe the expulsion of two vortices when the rf current is injected in parallel in the two STNOs with different diameters (200 and 400 nm). Note that, thanks to the circuit that has been developed, the measurements of the dc voltage drop observed when the external frequency equals the eigenfrequency of one of the two STNOs, is done separately on each STNO. A further advantage of having an individual electrical access to each STNOs is that, beyond the detection, we can also inject different dc currents in each STNOs, that is helpful to indeed control the spin-transfer forces acting on each vortex, meaning a way to control both the minimum threshold that can be detected as well as the frequency range in which the resistance level corresponds to a regime of vortex expulsion (see red region in the insert of Fig. 2c)). By this mean, we can thus tune the frequency bandwidth of the detector.

In conclusion, in this work we have focused on the expulsion of the vortex core, more adapted for the development of a threshold rf detector due to the better sensitivity compared to the spin-torque diode effect and the tunability of the expulsion frequencies. The connection of many STNOs with different diameters is crucial in order to cover the entire desired frequency band. We prove the concept of rf detection with two STNOs connected in parallel and measured separately. The next step will be to increase the number of connected STNOs and make the required optimization in the circuitry to achieve a system level demonstration of this kind of real-time nanoscale spectrum analyzer.

**Fig.1 – a)** Spin diode voltage as a function of the external frequency for a STNO of 400 nm diameter at $I_{dc}$ = 0 mA and $P_{rf}$ = -10 dBm. **b)** Expulsion of the vortex core measured with a pillar of 200 nm at $I_{dc}$ = 4 mA and $P_{rf}$ = -10 dBm with an external magnetic field of 500 Oe. **c)** Schematic plot of a vortex based STNO composed by two ferromagnetic layers separated by a non-magnetic layer.

**Fig.2 – a)** Measurement of vortices core expulsion, separately, for different diameters: 300, 400 and 500 nm. **b)** Dependence between the expulsion frequencies and the diameter of STNOs. **c,d)** Variation of expulsion (blue points) and nucleation (red points) as a function of **c)** the rf power at $I_{dc}$ = 10 mA with an in plane magnetic field H = 53.5 Oe and **d)** the in-plane magnetic field for a 400 nm diameter STNO at $I_{dc}$ = 9.8 mA and $P_{rf}$ = -10 dBm.

**Fig.3** – Measurement of two vortex core expulsions for two STNOs, connected in parallel and tuned with a dc current ($I_{dc}$ = 13 mA for D = 400 nm and $I_{dc}$ = 3 mA for D = 200 nm) and an in-plane magnetic field ($H_{in\_plane}$ = 57 Oe). The rf current is injected to both STNOs via a power divider.

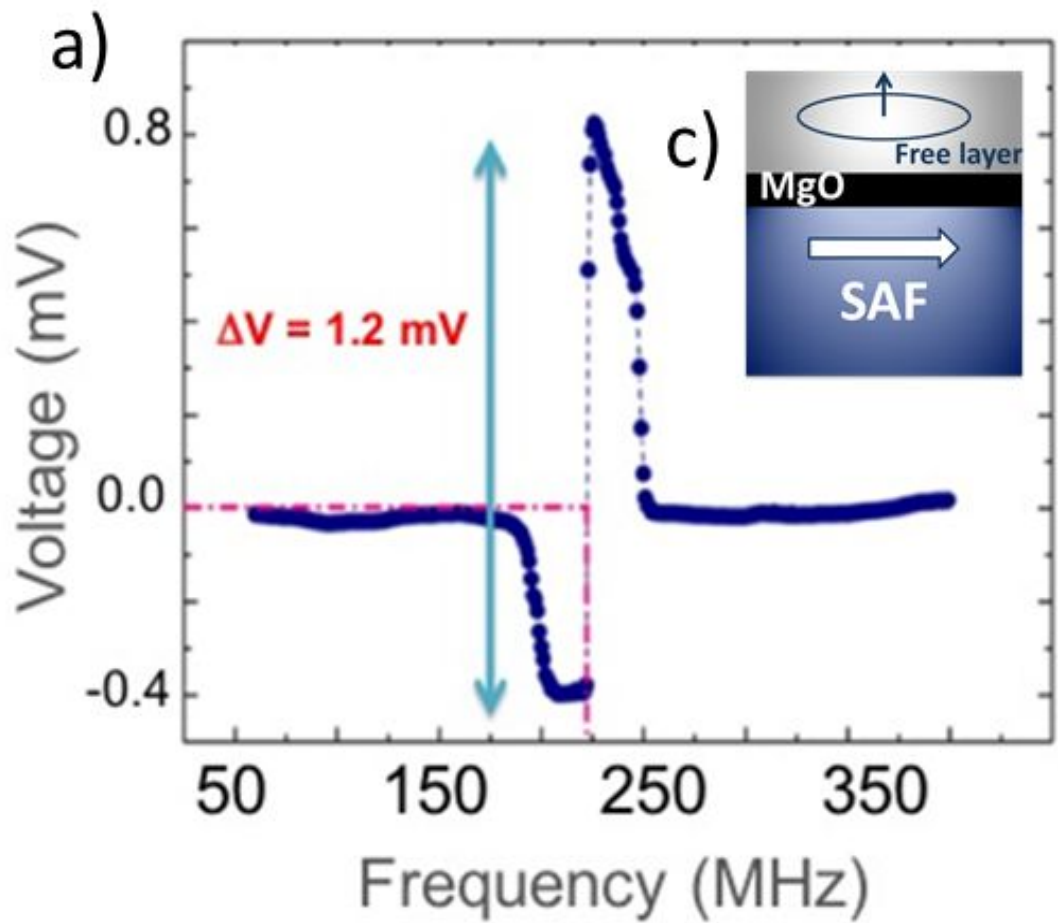
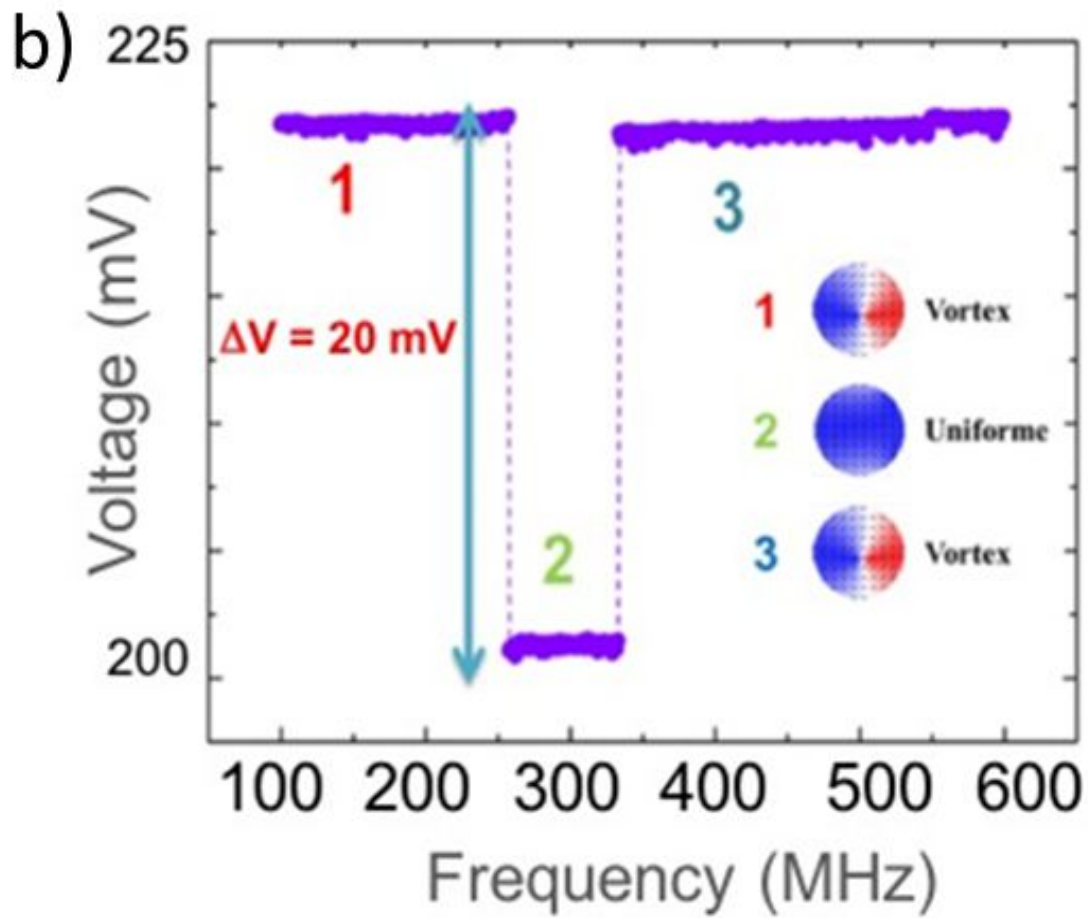

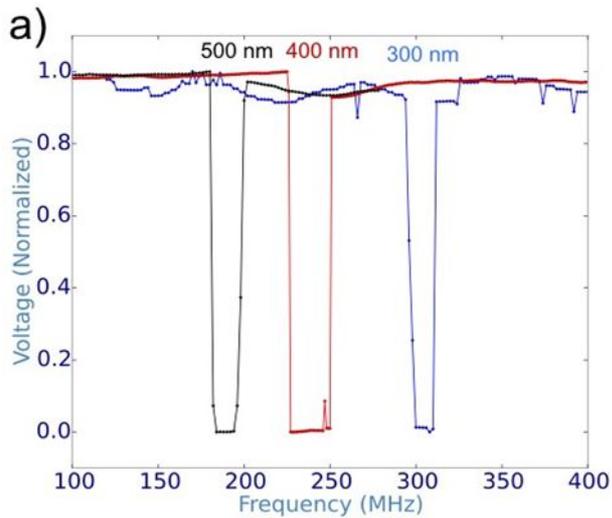
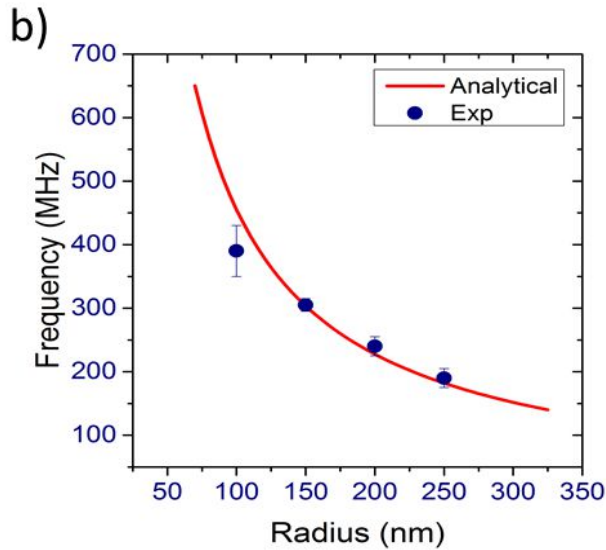
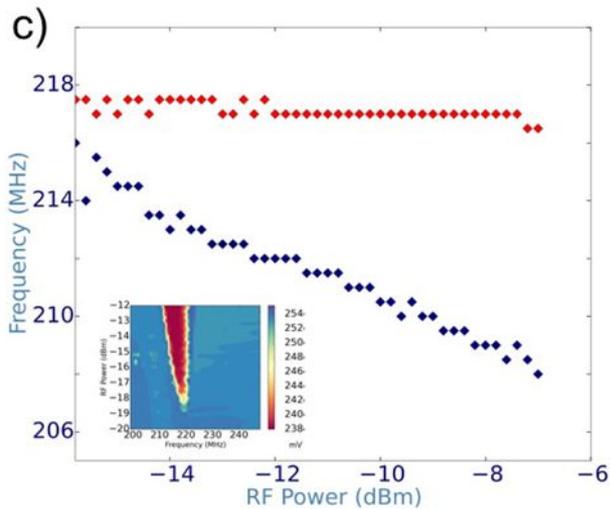
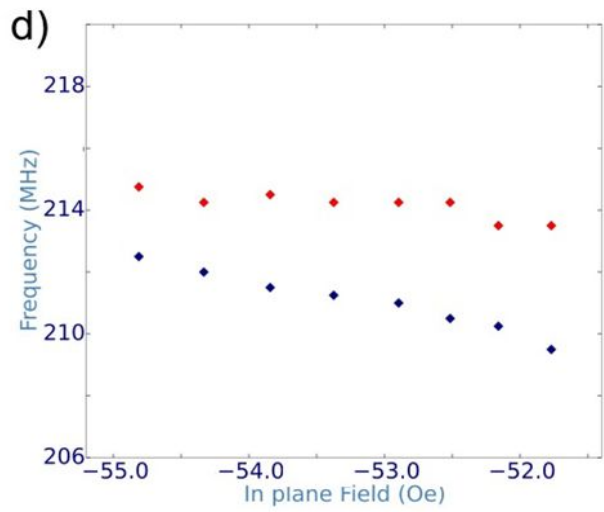

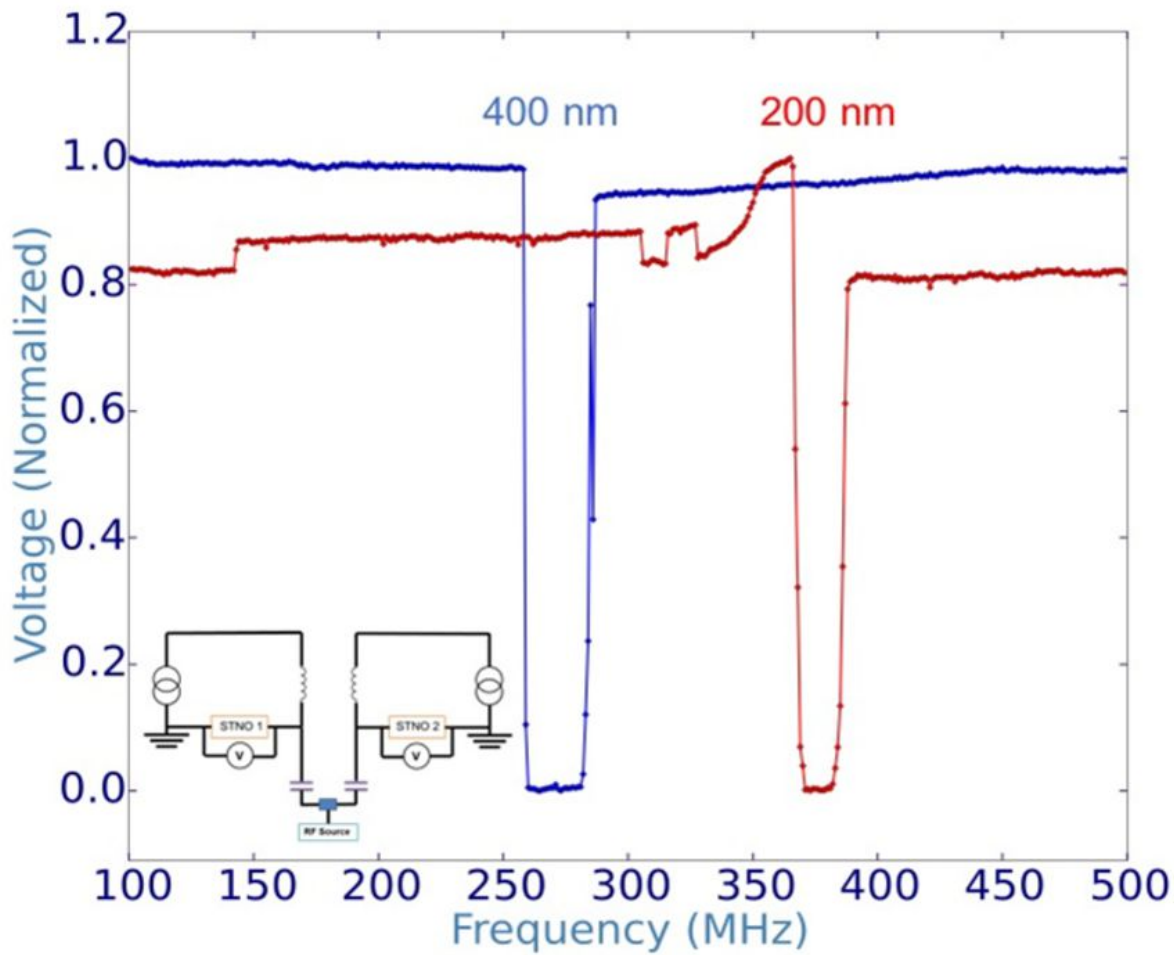